\documentclass[%
 aip,
 amsmath,amssymb,
 reprint,%
]{revtex4-1}

\usepackage{graphicx}
\usepackage{dcolumn}
\usepackage{bm}

\usepackage[utf8]{inputenc}
\usepackage[T1]{fontenc}
\usepackage{mathptmx}
\usepackage[colorlinks,linkcolor=blue,anchorcolor=blue,citecolor=blue,urlcolor=blue,CJKbookmarks=True]{hyperref}
\usepackage{changes}

\begin{document}

\preprint{AIP/123-QED}

\title{Reduced Ionic Diffusion by the Dynamic Electron-Ion Collisions in Warm Dense Hydrogen}

\author{Yunpeng Yao}
\author{Qiyu Zeng}
\author{Ke Chen}
\author{Dongdong Kang}
\author{Yong Hou}
\author{Qian Ma}
\author{Jiayu Dai}
\email{jydai@nudt.edu.cn}
\affiliation{Department of Physics, National University of Defense Technology, Changsha, Hunan 410073, P. R. China}

\date{\today}

\begin{abstract}
The dynamic electron-ion collisions play an important role in determining the static and transport properties of warm dense matter (WDM). Electron force field (eFF) method is applied to study the ionic transport properties of warm dense hydrogen. Compared with the results from quantum molecular dynamics and orbital-free molecular dynamics, the ionic diffusions are largely reduced by involving the dynamic collisions of electrons and ions. This physics is verified by the quantum Langevin molecular dynamics (QLMD) simulations, which includes electron-ion collisions induced friction (EI-CIF) into the dynamic equation of ions. Based on these new results, we proposed a model including the correction of collisions induced friction of the ionic diffusion. The CIF model has been verified to be valid at a wide range of density and temperature. We also compare the results with the Yukawa one component plasma (YOCP) model and Effective OCP (EOCP) model. We proposed to calculate the self-diffusion coefficients using the EOCP model modified by the CIF model to introduce the dynamic electron-ion collisions effect.
\end{abstract}

\maketitle

\section{\label{sec:one}INTRODUCTION}

Warm dense matter (WDM) is an intermediate state bridges the condesed matter and ideal plasma \cite{Fortov2011}. The transport properties of WDM, such as diffusion, viscosity, thermal conduction, and temperature relaxation, etc. \cite{Stanton2016,White2017,Heinonen2020,Collins2016,Barry2011}, play important roles in the field of astrophysics and inertial confinement fusion \cite{Nettelmann2008,Xu2011,DDKang2020,Rinderknecht2014,Huang2018} (ICF). For WDM, the ionic coupling parameter $\Gamma=Z_i^2e^2/\left(r_ik_BT\right)$ is larger than 1, and the electron degeneracy parameter $\Theta=T/T_F$ is less than 1. That requires us to consider both strong coupling between ions and the partial ionization and partial degeneration of electrons when studying the unique state. There has not been a mature theory that is good enough to describe the properties of WDM. At present, numerical simulation methods are more popular schemes, such as molecular dynamics (MD) methods \cite{Glosli2008,Haxhimali2014,QMa2014} and density functional theory \cite{JYDai2012,Kress2010,Horner2009} (DFT). Most of these models are based on the Born-Oppenheimer (BO) approximation. BO approximation---decouples ions from electrons to the instantaneously adjusting potential energy surface (PES) formed by fast electrons---has achieved great success on complex many-body systems. However, it may have difficulty in WDM considering the excitation and ionization of electrons. The drastic dynamic electron-ion collisions cause great disturbances in the PES, and the non-adiabatic effect will exhibit significant effects on the equilibrium and the non-equilibrium processes \cite{Andrew2012,Strickler2016,BBLu2019,QYZeng2020}. With the improvement of diagnostic methods, especially the application of X-ray Thomson scattering techniques \cite{Glenzer2009}, electronic information of WDM can be obtained in the laboratory. To interpret the experimental data, a more precise theory beyond BO approximation is required on account of the complex environment of WDM.  

The non-adiabatic effect has been considered by some methods to get more accurate interactions between electrons and ions in WDM. Derived from time-dependent Kohn-Sham equation, time-dependent density function theory \cite{Campetella2017} (TDDFT) gives the relatively exact electronic structure information. Thanks to the coupling of the electrons and ions, TDDFT-Ehrenfest approach can give the results such as energy dissipation process, excitation energies and optical properties \emph{etc} \cite{Graziani2014,Baczewski2016}. However, TDDFT is extremely time-consuming, limited by finite time and size scale. Thus low frequency modes can not be described well and the convergence of scale is required to be verified carefully. Quantum Langevin molecular dynamics (QLMD) holds a more efficient first principles computation efficiency, simultaneously regarding dynamic electron-ion collisions as frictional forces in Langevin dynamical equation of ions \cite{JYDai2010}. Using the QLMD method, a stronger ionic diffusive mode at low frequency has been found when the selected friction parameter becomes larger, as well as the decrease of the sound-speed \cite{Mabey2017}. Nevertheless, the determination of the friction parameter is \emph{a priori}. Recently, Simoni \emph{et al} have provided ab-initio calculations of the friction tensor in liquid metals and warm dense plasma \cite{Simoni2020}. They obtain a non-diagonal friction tensor, reflecting the anisotropy of instantaneous dynamic electron-ion collisions. Electron force field (EFF) expresses electrons as Gaussian wave packets, so that it can include the non-adiabatic effect intrinsically in molecular dynamics simulation \cite{Su2007,Jaramillo2011}. Lately the method has been applied to warm dense aluminum and found similar conclusions that non-adiabatic effect enhances ion modes around $\omega$ = 0, however, the effect is not sensitive to the sound speed \cite{Davis2020}. Q. Ma \emph{et al} have developed the EFF methodology to study warm and hot dense hydrogen \cite{QMa2019,QMa2018}. They conclude that dynamic electron-ion collisions reduce the electrical conductivities and increase the electron-ion temperature relaxation times compared with adiabatic and classical framing theories. As another approach, Bohmian trajectory formalism has been applied by Larder \emph{et al} recently \cite{Larder2019}. Constructing a thermally averaged, linearized Bohm potential, fast dynamical computation with coupled electronic-ionic system is achieved \cite{Larder2019}. The result also reveals different phenomenon of dynamic structure factor (DSF) and dispersion relation from DFT-MD simulation. All researches reflect that electron-ion collisions affect significantly on the study of dynamic properties of WDM, for both electrons and ions. Nevertheless, the effect of non-adiabatic effect on the ionic transport properties such as diffusion coefficient \cite{Hansen1975} is few studied, in both numerical simulations and analytical models. We could image the existence of dynamic electron-ion collisions will induce new effects such as dissipation or friction. In particular, for the analytical models based on traditional BO methods, we should study the non-adiabatic dynamic collisions effect on the self-diffusion in warm dense matter, and propose a new model including collisions induced friction (CIF).
 
The paper is organized as follows. Firstly, details of theoretical methods and the computation of diffusion coefficient are introduced in Section~\ref{sec:two}. Then, in Section~\ref{sec:three}, the static and transport results of QMD, OFMD, QLMD, and (C)EFF simulations are showed and the dynamic collisions effect is discussed. In section~\ref{sec:four}, we systematically study the colision frequency effect on ionic diffusions and the CIF model is introduced to estimate the impact of electron-ion collisions. In section~\ref{sec:five}, the results are compared with the YOCP, and EOCP models. Finally, the conclusions are given in section~\ref{sec:six}. All units are in atom unit if not emphasized. 

\section{\label{sec:two}THEORETICAL METHODS AND COMPUTATIONAL DETAILS}

\subsection{(Constrained) electron force field methodology}

EFF method is supposed to be originated from wave packet molecular dynamics (WPMD) \cite{Heller1975} and floating spherical Gaussian orbital (FSGO) method \cite{Frost1967}. Considering each electronic wave function as a Gaussian wave packet, the excitation of electrons can be included with the evolution of positions and wavepacket radius. N-electrons wave functions are taken as a Hartree product of single-electron Gaussian packet written as
\begin{eqnarray}
     \Psi\left({\bf r}\right)&=&\left(\frac{2}{s^{2}\pi}\right)^{3/4}\exp\left(-\left(\frac{1}{s^2}-\frac{2p_{s}i}{s\hbar}\right)\left({\bf r}-{\bf x}\right)^{2}\right)\nonumber \\ & &\cdot\exp\left(\frac{i}{\hbar}{\bf p_x}\cdot{\bf r}\right).
\end{eqnarray}
where $s$ and ${\bf x}$ are the radius and average positions of the electron wave packet, respectively. $p_s$ and ${\bf p_x}$ correspond to the conjugate radial and translational momenta. Nuclei in EFF are treated as classical charged particles moving in the mean field formed by electrons and other ions.
 
Substituting simplified electronic wave function in the time-dependent Schr\"odinger equation with a harmonic potential, equation of motion for the wave packet can be derived
 \begin{subequations}
  \begin{equation}
     {\bf \dot{x}}={\bf p_x}/m_e, 
  \end{equation}
  \begin{equation}
     {\bf \dot{p_x}}=-\nabla_{x}V, 
   \end{equation}
   \begin{equation}
     \dot{s}=(4/d)p_{s}/m_e,  
   \end{equation}
    \begin{equation}
     \dot{p_s}=-\partial{V}/\partial{s}.
   \end{equation}
 \end{subequations}
where $d$ is the dimensionality of wave packets. For a three-dimensional system, $d$ is equal to 3, and it becomes 2 in 2D systems. $V$ is the effective potential. Combining with ionic equations of motion, EFF MD simulations have been implemented in LAMMPS package\cite{Jaramillo2011}.  
 
In addition to electrostatic interactions and electron kinetic energy, spin-dependent Pauli repulsion potential is added in the Hamiltonian as the anti-symmetry compensation of electronic wave functions. In EFF methodology, the exchange effect is dominated by kinetic energy. All interaction potentials are expressed respectively as
 \begin{subequations}
 \begin{equation}
     E_{nuc-nuc}=\sum_{i<j} Z_{i}Z_{j}/R_{ij},
 \end{equation}
  \begin{equation}
     E_{nuc-elec}=\sum_{i<j} -\left(Z_{i}/R_{ij}\right)\emph{erf}\left(\sqrt{2}R_{ij}/s_j\right),
 \end{equation}
 \begin{equation}
     E_{elec-elec}=\sum_{i<j} \left(1/r_{ij}\right)\emph{erf}\left(\sqrt{2}r_{ij}/\sqrt{s_{i}^{2}+s_{j}^{2}}\right),
 \end{equation}
 \begin{equation}
     E_{ke}=\sum_{i} \left(3/2\right)\left(1/s_{i}^2\right),
 \end{equation}
  \begin{equation}
     E_{Pauli}=\sum_{\sigma_{i}=\sigma_{i}} E\left(\uparrow\uparrow\right)_{ij}+\sum_{\sigma_{i}\neq\sigma_{i}} E\left(\uparrow\downarrow\right)_{ij}.
 \end{equation}
 \end{subequations}
where $Z$ is the charge of nucleus, $r_{ij}$ and $R_{ij}$ correspond to the relative positions of two particles (nuclei or electrons). $\emph{erf}\left(x\right)$ is error function, $\sigma$ means the spin of electrons. Pauli potential is consists of same and opposite spin electrons repulsive potentials. More details can be found in Refs.~\onlinecite{Su2007,Jaramillo2011,Su2009,Patrick2012}.
 
However, the EFF model also suffers from the limitation of WPMD. That is the wave packets spread at high temperature \cite{Grabowski2013}. To avoid excessive spreading of wave packets, the harmonic constraints are often added. Recently, Constrained EFF (CEFF) method has been proposed using $L=\lambda_{D}+b_0$ as the boundary of the wave packets \cite{QMa2019}, getting much lower electron-ion energy exchange rate agreeing with experimental data \cite{Celliers1992,White2014}. 

We use the EFF method to calculate the static and transport properties of hydrogen at $\rho=5\text{g/cm}^3$ and $\rho=10\text{g/cm}^3$. The temperature is from 50kK to 300kK. In the simulations, the real electron mass is used so that we choose the time step as small as $0.2\emph{as}$. 1000 ions and 1000 electrons are used in the simulation. $5\emph{ps}$ microcanonical ensemble with a fixed energy, volume, and number of particles (NVE) has been performed to calculate statistical average after $10\emph{ps}$ simulations with fixed temperature, volume, and number of particles (NVT). When the temperature becomes higher, CEFF is applied to avoid wave packets spreading \cite{QMa2019}.

\subsection{Quantum molecular dynamics and orbital-free molecular dynamics}

For comparision, we also run the adiabatic simulations including QMD and OFMD. The QMD simulations have been performed using Quantum-Espresso (QE) open-source software \cite{Giannozzi2009}. In QMD simulations, electrons are treated quantum mechanically through the finite temperature DFT (FT-DFT). While ions evolve classically along the PES determined by the electric density, and the electron-ion interaction is described as plane wave pseudopotential. Each electronic wave function is solved by the Kohn-Sham equation \cite{Cowan2001}
\begin{equation}
     \left(-\frac{1}{2}\nabla^{2}+V_\text{KS}[n_e\left({\bf r}\right)]\right)\varphi_i\left({\bf r}\right)=E_i\varphi_i\left({\bf r}\right).
\end{equation}
where $E_i$ is the eigenenergy, $-1/2\nabla^2$ is the kinetic energy contribution, and the Kohn-Sham potential $V_\text{KS}[n_e\left({\bf r}\right)]$ is given by
\begin{equation}
    V_\text{KS}[n_e\left({\bf r}\right)]=\upsilon\left({\bf r}\right)+\int \frac{n_{e}\left({\bf r}'\right)}{|{\bf r}-\bf{r}'|}d{\bf r}'+V_{\text{xc}}[n_e\left({\bf r}\right)].
\end{equation}
where $\upsilon\left({\bf r}\right)$ is the electron-ion interaction, the second term in the right hand of the above equation is the Hartree contribution, an $V_{\text{xc}}[n_e\left({\bf r}\right)]$ represents the exchange-correlation potential, which is represented by the Perdew-Burke-Ernzerhof (PBE) functional \cite{Perdew1996} in the generalized-gradient approximation (GGA) during the simulations. The electronic density consists of single electronic wave function
\begin{equation}
     n_{e}\left({\bf r}\right)=2\sum_{i}\left|\varphi_i\left({\bf r}\right)\right|^2.
\end{equation}

In our simulations, only the $\Gamma$ point $\left(\boldsymbol{k}=0\right)$ is sampled in the Brillouin-zone, and we used supercells containing 256 H atoms. The velocity Verlet algorithm \cite{Verlet1967} is used to update position and velocity of ions. The time step is set from $0.05\emph{fs}$ to $0.1\emph{fs}$ at different temperature to ensure convergence of energy. The cutoff energy is tested and set from 100 Ry to 150 Ry. The number of bands is sufficient for the occupation of electrons. Each density and temperature point is performed for at least 4000-10000 time steps in the canonical ensemble, and the ensemble information is picked up after the system reaches equilibrium.
 
At high temperatures, the requirement of too many bands limits the efficiency of QMD method. OFMD is a good choice when dealing with high temperature conditions \cite{ZQWang2020,Clerouin1997,Collins1995,Meyer2014}. Within orbital-free frame, the electronic free energy is expressed as
\begin{eqnarray} 
    & & F_{e}{\bf [}{\bf R},n_e{\bf ]}= \frac{1}{\beta}\int d{\bf r}\Big{\{}{n_e\left({\bf r}\right)\Phi[n_e\left({\bf r}\right)]-\frac{2\sqrt{2}}{3\pi^2\beta^{3/2}}I_{\frac{3}{2}}\{{\Phi[n_e({\bf r})]}\}}\Big{\}} \nonumber \\ & & +\int d{\bf r}V_{\text{ext}}\left({\bf r}\right)+\frac{1}{2}\iint d{\bf r}d{\bf r}'\frac{n_e\left({\bf r}\right)n_e\left({\bf r}'\right)}{|{\bf r}-{\bf r}'|}+F_{\text{xc}}[n_e\left({\bf r}\right)].
\label{eq:freenergy}
\end{eqnarray} 
where ${\bf R}$ is the ionic position, $\beta=1/k_BT$ where $T$ is the temperature and $k_B$ is the Boltzmann constant. $I_\nu$ is the Fermi integral of order $\nu$. $V_{\text{ext}}\left({\bf r}\right)$ represents the external or the electron-ion interaction, and $F_{\text{xc}}[n_e\left({\bf r}\right)]$ is the exchange-correlation potential. The electrostatic screening potential is represented by $\Phi[n_e({\bf r})]$ depend on electronic density $n_e({\bf r})$ only
\begin{equation}
     \nabla^2\Phi[n_e\left({\bf r}\right)]=4{\pi}n_e\left({\bf r}\right)=\frac{4\sqrt{2}}{\pi^2\beta^{3/2}}I_{\frac{1}{2}}\{\Phi[n_e\left({\bf r}\right)]\}.
\end{equation}

The OFMD simulations are performed with our locally modified version of PROFESS\cite{MHChen2015}. The PBE functional \cite{Perdew1996} is also used to treat the exchange-correlation potential. 256 H atoms are also used in the supercell. The kinetic energy cutoff is $7000eV$ when the density $\rho=5\text{g/cm}^3$, and $10000eV$ at $\rho=10\text{g/cm}^3$. The time step is set from $0.04\emph{fs}$ to $0.15\emph{fs}$ with the temperature increasing. The size effect has been tested in all MD simulations.
 
\subsection{Quantum Langevin molecular dynamics}

QMD and OFMD are good tools in describing static properties of warm dense matters. However, the information of electron-ion dynamical collisions is lost because of the assumption of BO approximation. In addition, electron-ion collisions are important for WDM in which electrons are excited because of the increasing temperature and density. To describe the dynamic process, QMD has been extended by considering the electron-ion collision induced friction (EI-CIF) in Langevin equation, and corresponding to the QLMD model \cite{JYDai2010}. In QLMD, ionic trajectory is performed using the Langevin equation\cite{DDKang2018}.
 \begin{equation}
     M_{I}\ddot{{\bf R}}_I={\bf F}-{\gamma}M_I\dot{{\bf R}}_I+{\bf N}_I.
 \end{equation} 
 where $M_I$ and ${\bf R}_I$ is the mass and position of the ion respectively, ${\bf F}$ is the force calculated from DFT simulation, $\gamma$ means the friction coefficient, and ${\bf N}_I$ represents a Gaussian random noise. In QLMD, the force produced by real dynamics of electron-ion collisions can be replaced by the friction on account of less time scale of electronic motions comparing with that of ions. The friction coefficient $\gamma$ is the key parameter should be determined \emph{a-priori}. Generally, at high temperature such as WDM and HDM regimes, the EI-CIF dominates the friction coefficient, and can be estimated from the Rayleigh model \cite{Plyukhin2008} 
\begin{equation}
     \gamma=2\pi\frac{m_e}{M_I}Z^*(\frac{4{\pi}n_i}{3})^{1/3}\sqrt{\frac{k_BT}{m_e}}.
     \label{eq:Rayliegh}
\end{equation} 
where $m_e$ is the electronic mass, $Z^*$ is the average ionization degree. In the paper, we used average atom (AA) model, in which the energy level broadening effect is considered, to estimate the average ionization degree. $n_i$ means the ionic number density. There is another way to assess $\gamma$ based on the Skupsky model\cite{Skupsky1977,Stanton2018}, and in this work, we adopted Rayleigh model only considering the hydrogen we studied has high density and high temperature.
 
To make sure that the particle velocity satisfies the Boltzmann distribution, the Gaussian random noise ${\bf N}_I$ should obey the fluctuation-dissipation theorem \cite{JYDai2009}
 \begin{equation}
     \left \langle {\bf N}_I \left(0\right){\bf N}_I\left(t\right) \right \rangle=6{\gamma}M_Ik_BTdt.
 \end{equation} 
where $dt$ is the time step in the MD simulation. The angle bracket denotes the ensemble average.

\subsection{Self-diffusion coefficient}

In MD simulations, the self-diffusion coefficient is often calculated from the velocity autocorrelation function (VACF) using Green-Kubo formula \cite{Kubo1957}
 \begin{subequations}
 \begin{equation}
     D=\lim_{t \to \infty}D\left(t\right),
 \end{equation}
 \begin{equation}
     D\left(t\right)=\frac{1}{3}\int_{0}^{t}dt\left \langle 
     {\bf v}_i\left(t\right)\cdot {\bf v}_i\left(0\right)\right \rangle.
 \end{equation}
 \end{subequations}
in which ${\bf v}_i\left(t\right)$ is the center of mass velocity of the $i$th particle at time $t$, and the angle bracket represents the ensemble average. Generally, the integral is computed in long enough MD trajectories so that the VACF becomes nearly zero and has less contribution to the integral. All the same species of particles are considered in the average to get faster convergent statistical results.
 
In practical, it is impossible to get a strict convergent result because the infinite simulation is forbidden. Thus, we usually use a exponential function $\left \langle {\bf v}\left(t\right)\cdot {\bf v}\left(0\right)\right \rangle=a\exp {\left(-t/\tau\right)}$ to fit the VACF to get the self-diffusion coefficient $D=a\cdot \tau$. Where $a$ and $\tau$ are fitting parameters determined by a least-squares fit. $\tau$ is corresponds to the decay time. In moderate and strong coupling regimes, a more sophisticated fitting expression is need to be considered \cite{Meyer2014}. In the exponential function fitting, the statistical error can be estimated by \cite{Hess2002}
 \begin{equation}
     \epsilon=\sqrt{\frac{2\tau}{NT_{\emph{traj}}}}
 \end{equation}
where $N$ is the number of particles, $T_{\emph{traj}}$ is the total time in the MD simulation. 

\section{\label{sec:three}RESULTS AND DISCUSSION}
\subsection{Static and transport properties}
 
\begin{figure}
     \centering
           \includegraphics[width=0.4\textwidth]{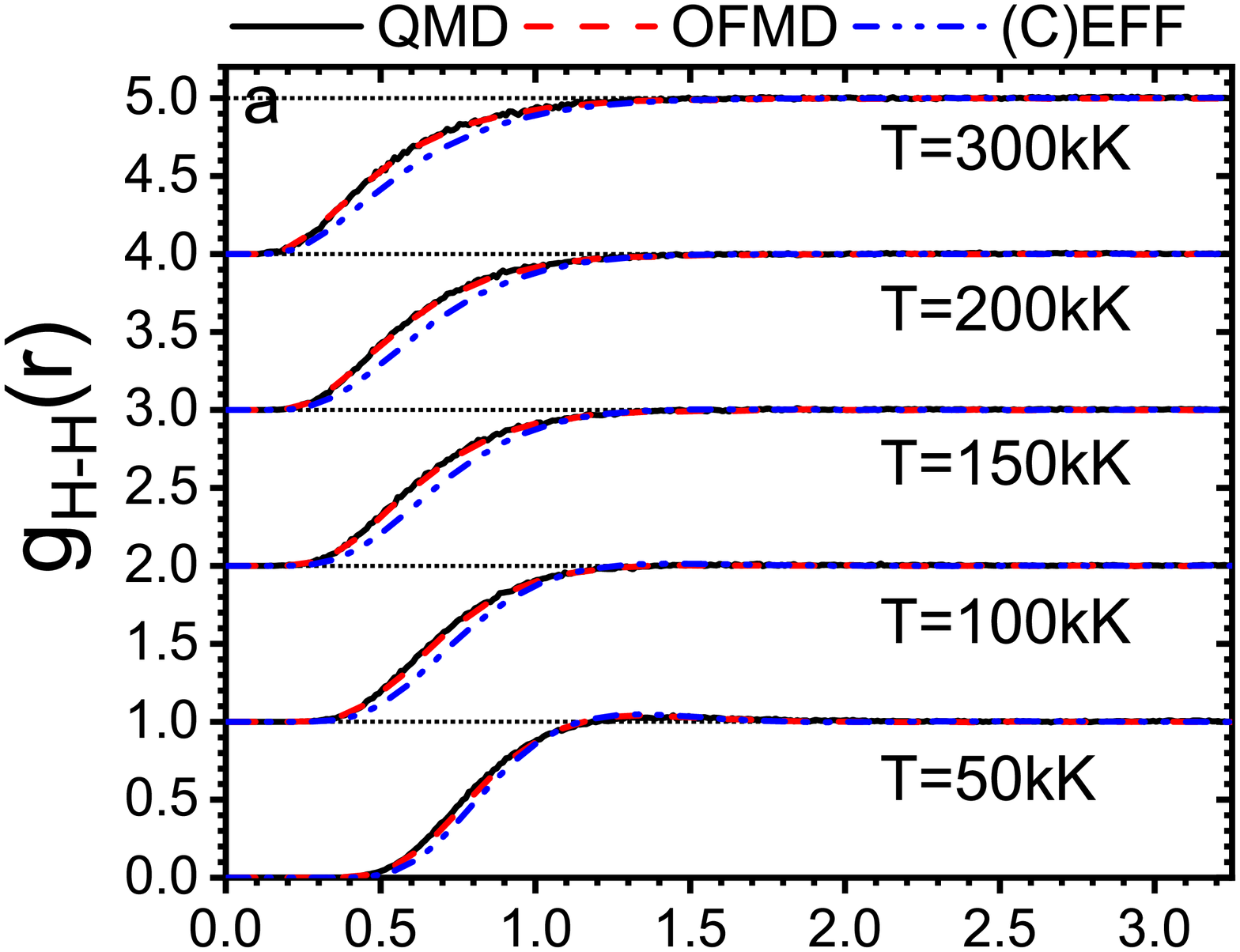} \\
           \includegraphics[width=0.4\textwidth]{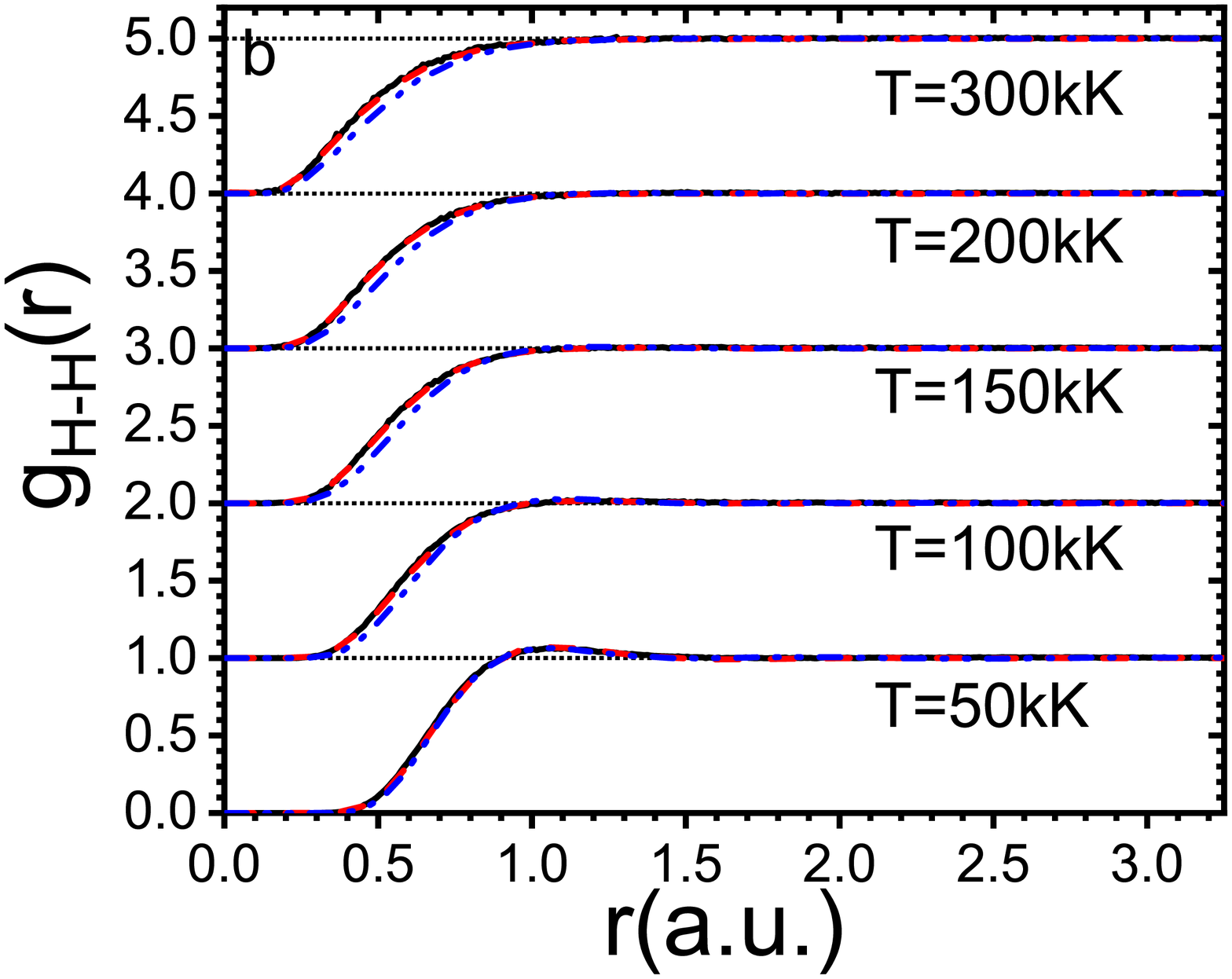}
     \caption{\label{fig:rdf}The RDFs of H-H at $5\text{g/cm}^3$ (a) and $10\text{g/cm}^3$ (b). The ordinate is differentiated by adding factors at different temperatures. Blue double dots lines represent the results from (C)EFF simulation. Black solid and red dashed lines are the QMD and OFMD results, respectively.}
 \end{figure}

We firstly calculate the radial distribution function (RDF) $g\left(r\right)$ of H-H, as shown in Fig.~\ref{fig:rdf}. It is shown that the RDFs from OFMD calculations agree well with RDFs of QMD results. Moreover, the RDFs calculated from (C)EFF reflect similar microscopic characteristics with QMD and OFMD results, especially when the temperature is relatively low, where the electron-ion collisions are not so important. For these cases, it is appropriate to show the intrinsic different physics between static and transport properties if the RDFs shown are very close to each other. It should be noticed that the RDFs of (C)EFF model shows a little more gradual than QMD's and OFMD's with the increase of temperature. It is deduced that the non-adiabatic effect plays little role in the static structures of warm dense hydrogen shown here, which is similar to the effects of Langevin dynamics on the static structures \cite{JYDai2009,Mabey2017}, in which the choice of friction coefficients has little effect on the RDFs.

\begin{figure}
     \centering
           \includegraphics[width=0.4\textwidth]{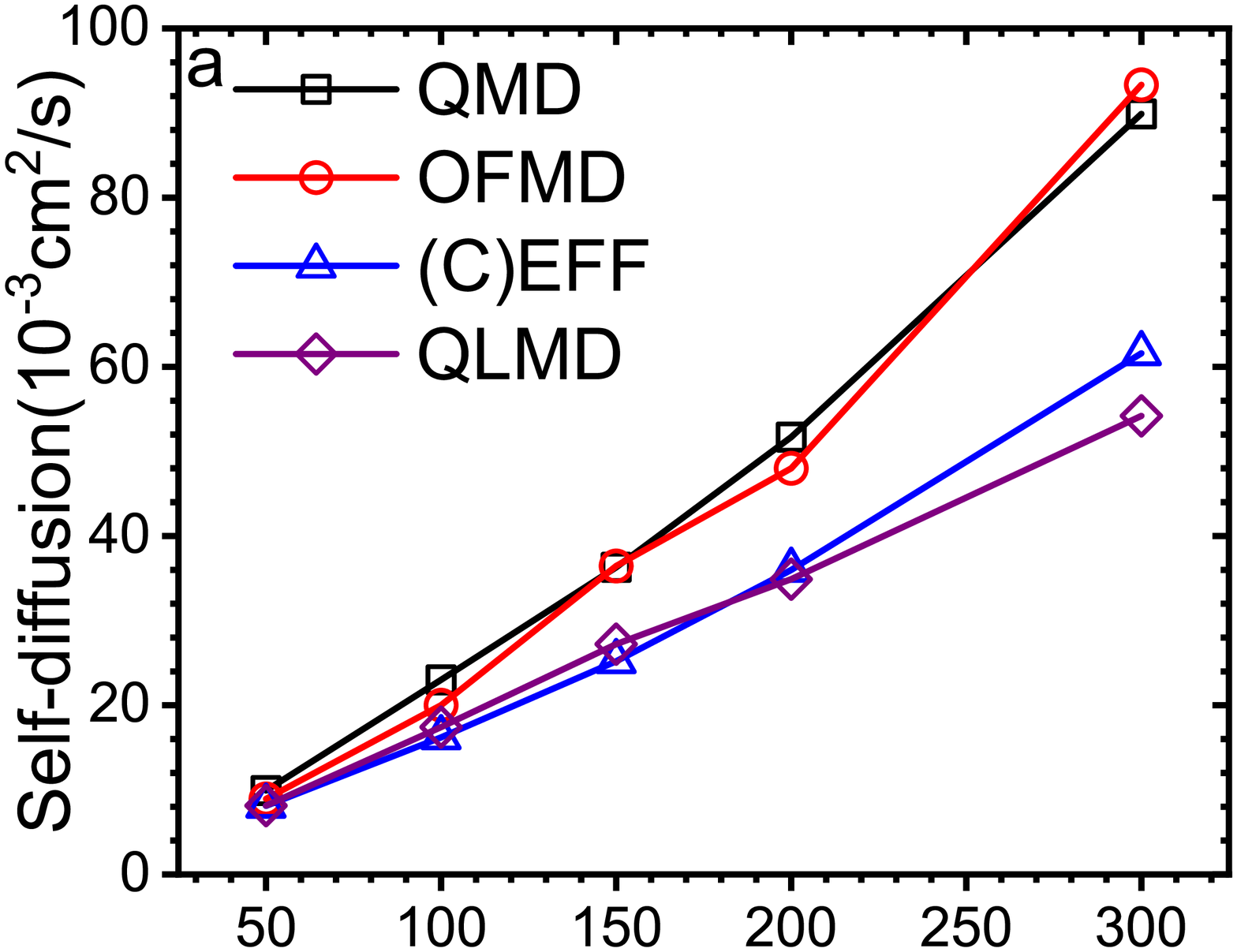}   \\
           \includegraphics[width=0.4\textwidth]{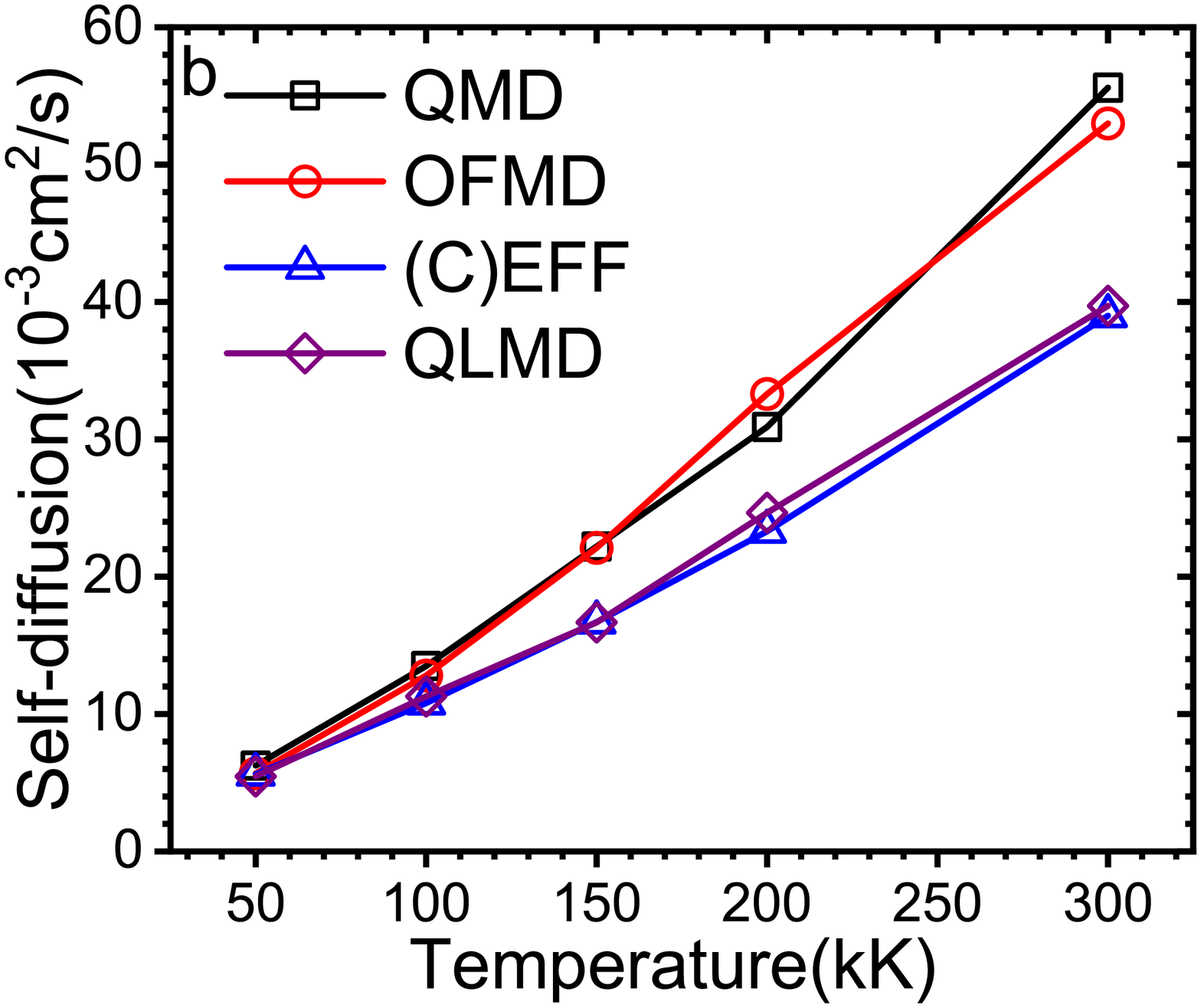}
     \caption{\label{fig:diff_simu}The self-diffusion coefficients of H as a function of temperature at $5\text{g/cm}^3$ (a) and $10\text{g/cm}^3$ (b), calculated by QMD, OFMD, QLMD, and (C)EFF methods. Black squares represent the QMD results, red circles are the results of OFMD's, and the (C)EFF results are represented by blue triangles. The QLMD results\cite{JYDai2010} are represented by purple diamonds}
 \end{figure}

However, non-adiabatic effects on dynamic properties are significant \cite{Mabey2017,JYDai2010,Larder2019}. We calculated the self-diffusion coefficients for warm dense hydrogen by integrating the VACF. To get a convergent value, a simple exponential function mentioned in Section~\ref{sec:two} is applied. The self-diffusion coefficient varies with temperature at $5\text{g/cm}^3$ and $10\text{g/cm}^3$ are shown in Fig.~\ref{fig:diff_simu} using different methods of (C)EFF, QMD, and OFMD.

It is very interesting that three methods give consistent results when temperature is relatively low. And the OFMD and QMD results have close values even with the increase of temperature. However, the (C)EFF simulations have a distinct reduce on the self-diffusion coefficients comparing with QMD and OFMD results. And the difference becomes more obvious at higher temperature. We boil it down to the non-adiabatic electron-ion dynamic collisions, which is lost in the framework of BO approximation such as QMD and OFMD. Regarding the electron as a Gaussian wave packet, (C)EFF methodology implements the electron-ion dynamics simulations, in which the dynamic coupling and collisions can be naturally included. As shown in Fig.~\ref{fig:diff_simu}, with the temperature increase, more electrons are excited or ionized and become free electrons. These free electrons lead continual and non-negligible electron-ion collisions, supplying drag forces for the motion of ions, and giving rise to much lower diffusion coefficients. The collision rate increases with the temperature, showing lower diffusive properties for ions, significantly affects the transport properties of WDM.

The lost of dynamic collisions can be introduced into the QMD model by considering electron-ion collision induced friction in Langevin equation. Here, we use the Rayliegh model to estimate the friction coefficient $\gamma$, and the QLMD simulations have been performed. It is very exciting that the QLMD results, showed in Fig.~\ref{fig:diff_simu}, agree well with (C)EFF simulations. The greatest difference between the two models is 12\%, but mostly within 6\%. This suggests that the reduction in ionic diffusion from (C)EFF simulations does indeed come from electron-ion dynamic collisions. We believe the small difference belongs to the choice of friction coefficient $\gamma$. Since the prior parameter should be determined artificially in QLMD simulations, we are encouraged to do quantitative analysis about the electron-ion collisions effect using the (C)EFF results as benchmark for the results of all adiabatic methods and analytical models.
 
 \section{\label{sec:four}ELECTRON-ION COLLISIONS EFFECT ASSESSMENT}

As shown above, we should figure out the mechanism how does the dynamic collisions work on the ionic transport? We can find a clue from the Landau-Spitzer (LS) electron-ion relaxation rate $\left(\nu_{ei}\right)$ \cite{Landau1937,Spitzer1967}
 \begin{equation} 
     \label{eq:electron-ioncollision}
     \nu_{ei}=\frac{8\sqrt{2\pi}n_iZ^2e^4}{3m_em_i}\left(\frac{k_BT_e}{m_e}+\frac{k_BT_i}{m_i}\right)^{-3/2}\ln{\Lambda}
 \end{equation} 
where $m_e\left(m_i\right)$, $n_e\left(n_i\right)$ and $T_e\left(T_i\right)$ are the mass, number density and temperature of electrons(ions), respectively. The Coulomb logarithm $\ln{\Lambda}$ can be calculated by the GMS model \cite{Gericke2002}. In Eq. \ref{eq:electron-ioncollision}, it is obvious that with the increase of density and temperature (the Coulomb logarithm also varies with the density and temperature), the collision frequency becomes higher, leading the diffusion coefficients reduce more significantly. The results of QMD and (C)EFF simulations showed in Fig.~\ref{fig:diff_simu} exhibit the same behaviors.

\begin{figure}
     \centering
     \includegraphics[width=0.4\textwidth]{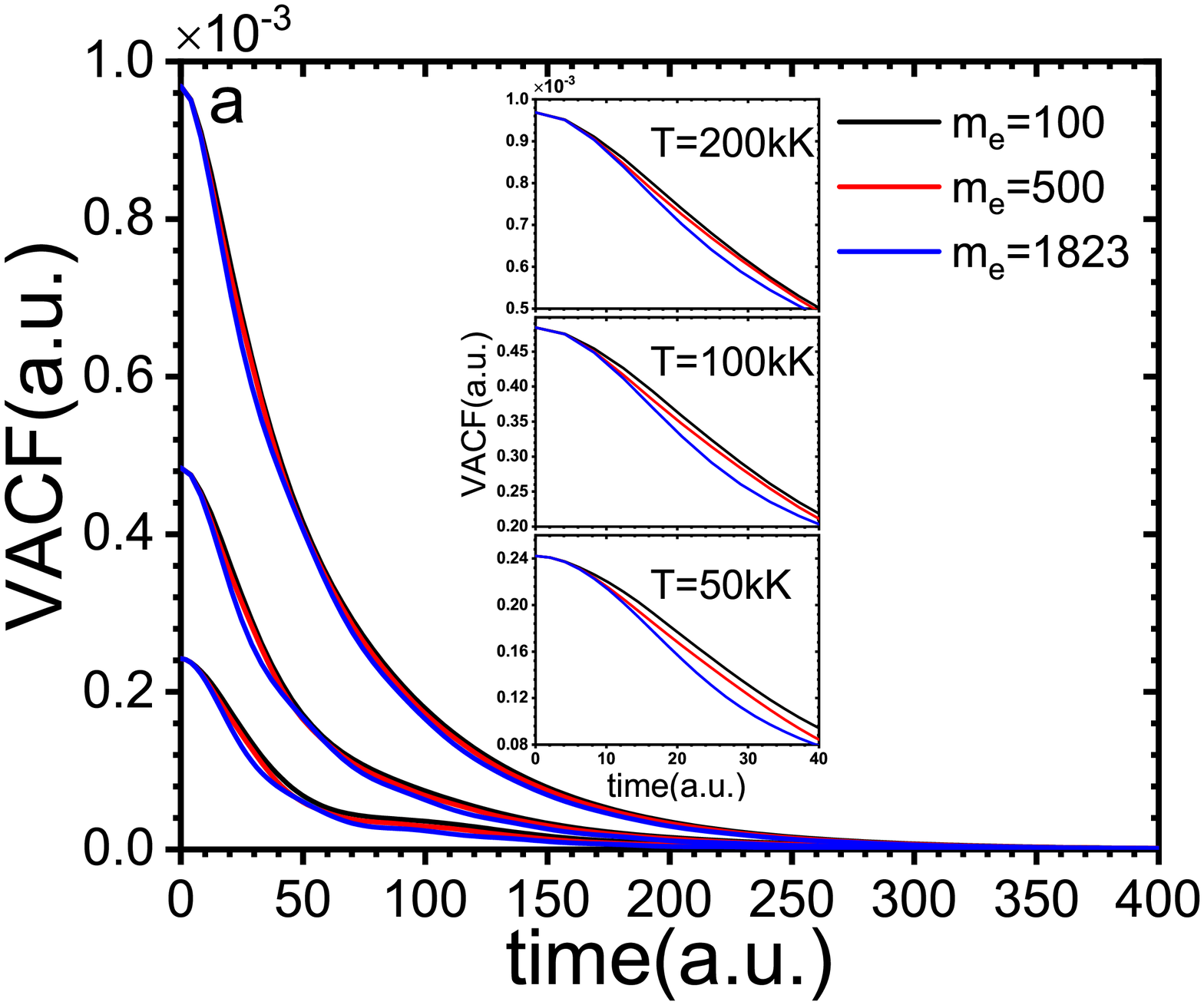}  \\
     \includegraphics[width=0.4\textwidth]{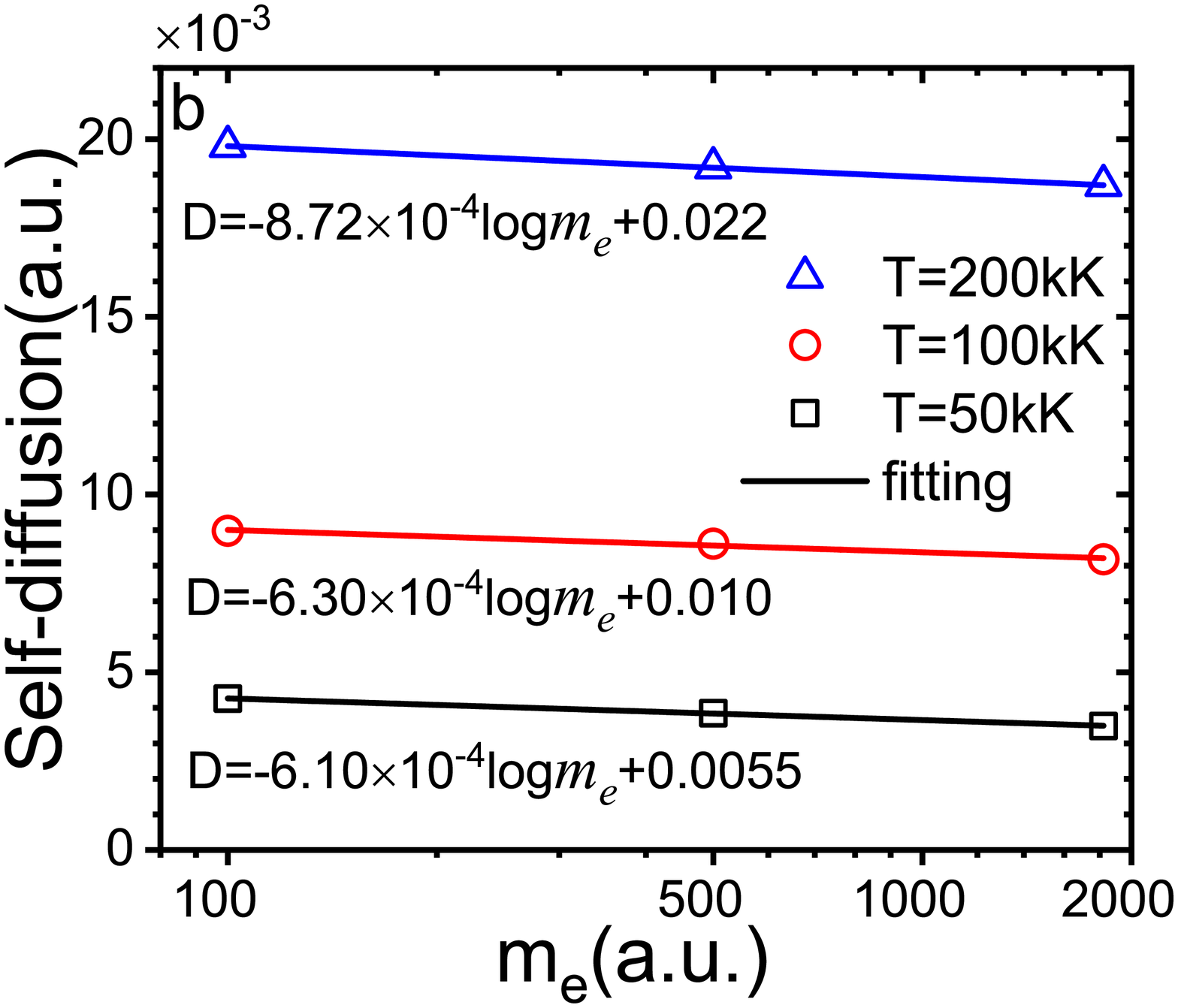}
     \caption{\label{fig:masseffect}(a)  VACFs of H for different dynamic electron mass at 200kK, 100kK, and 50kK from top to bottom. The density is $10\text{g/cm}^3$. The black lines, red lines, and blue lines represent the dynamic electron mass of 100a.u., 500a.u., and 1823a.u., respectively. Details are showed in the insets. (b) The corresponding ionic self-diffusion coefficients as a function of dynamic electron mass. The squares, circles, and triangles represent the temperature at 50kK, 100kK, and 200kK, respectively. The lines are the fitting results, and the fitting functions are listed below the lines.}
 \end{figure}

As shown in Eq.~\ref{eq:electron-ioncollision}, the electron-ion relaxation rate ($\nu_{ei}$) is the function of temperature and density. However, the thermodynamic state also changes with temperature and density, therefore it is difficult to distinguish the electron-ion collisions effect. For this purpose, we can change the effective mass of electrons in the (C)EFF simulation without altering the intrinsic interactions in the Hamiltonian of the system\cite{Jaramillo2011,Theofanis2012}. Since the mass of ions is much greater than that of electrons, we can find a simple relationship between the electron-ion collision frequency $\nu_{ei}$ and the mass of the electron $m_e$ from Eq.~\ref{eq:electron-ioncollision}
 \begin{equation} 
     \label{eq:nuvsme}
     \nu_{ei}=f\left(\rho,T\right)m_e^{1/2}
 \end{equation}      
When the dynamic electron mass is larger, the motion of effective electrons exhibit more classical, and the collisions between electrons and ions become stronger. By this way, we can study the influence of electron-ion collisions by adjusting electronic mass in (C)EFF simulations. The VACFs and self-diffusion coefficients of H at different dynamic electron mass are showed in Fig.~\ref{fig:masseffect}.
 
\begin{figure}
     \centering
     \includegraphics[width=0.45\textwidth]{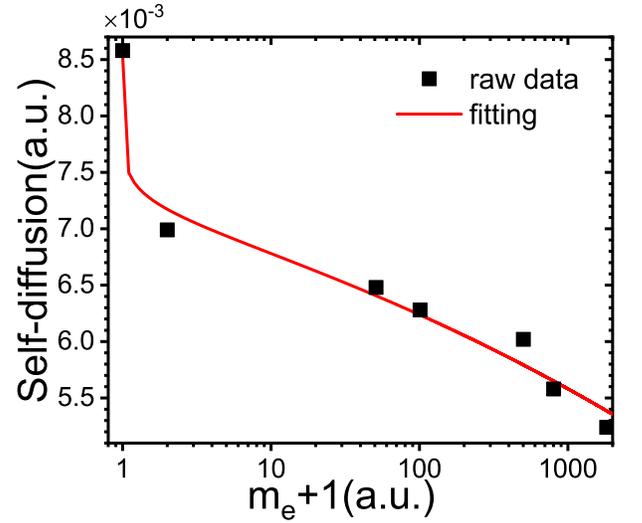}
     \caption{\label{fig:massmodel}Dynamic electron mass effects on ionic diffusion. We show the (C)EFF simulation results with different dynamic electron mass at $5\text{g/cm}^3$ and the temperature is 5kK. The mass of electrons has been shifted to avoid infinity definition of log function at zero point. The value at zero point is replaced by the QMD result. The fitting result is represented by the red line.}
 \end{figure}

From the VACF results we can see, The change of dynamic electron mass does not alter the thermodynamic states of ions. While, dynamic collisions reduce the correlation of particles, showing lower decay time with the increase of dynamic electron mass, as well as the electron-ion collision frequency. Diffusions reflect similar trends, and more interestingly, the diffusion of ions is inversely proportional to the log of electronic mass as showed in Fig.~\ref{fig:masseffect}(b). The inverse ratio relation reflects the reduction of the diffusion due to electron-ion collisions, and the slope determines the magnitude of this influence. In Fig.~\ref{fig:masseffect}(b), it is shown that the influence of electron-ion collisions becomes stronger with the increase of temperature, since the electrons are more classical at higher temperature. To quantitatively describe the relationship between diffusion and collision frequency, we performed more intensive simulations on dynamic electron mass as showed in Fig.~\ref{fig:massmodel}. Here, the QMD results are used as the value at reference point, corresponding to no dynamic electron-ion collisions, since $m_e$ can not be zero.

As showed in Fig.~\ref{fig:massmodel}, the change of diffusion coefficients decreases much steeper when the electron dynamic mass becomes smaller, revealing more significant effect of electron-ion collisions. Another decaying function as $D=a\log{\left(1+bm_e^c\right)}+d$ can well describe this relation of diffusion varying with the dynamic electron mass $m_e$. This function can transit to the linear form when $m_e$ is large. Here, we have found the approximate relationship between ionic diffusion coefficient $D$ and electron-ion collision rate $\nu_{ei}$ taking Eq.~\ref{eq:nuvsme} into the fitting function
 \begin{equation} \label{eq:fittingfunction}
     D=f_1\left(\rho,T\right)\log{\left(1+f_2\left(\rho,T\right)\nu_{ei}^{f_3\left(\rho,T\right)}\right)}+f_4\left(\rho,T\right)
 \end{equation}
where $f_1\left(\rho,T\right),f_2\left(\rho,T\right),f_3\left(\rho,T\right),f_4\left(\rho,T\right)$ are the function of the density $\rho$ and temperature $T$. If $\nu_{ei}$ is set to zero, the first term in the right hand of Eq.~\ref{eq:fittingfunction} vanishes, and $D=f_3\left(\rho,T\right)=D_0$. Here, the remaining term $D_0$ represents the diffusion without electron-ion collisions. We call the first term as collisions induced friction (CIF) of the ionic diffusion $D_{\text{CIF}}$. Within this consideration, the total diffusion coefficient can be obtained via
 \begin{equation} \begin{split} \label{eq:diffsplit}
     D&=f_1\left(\rho,T\right)\log{\left(1+f_2\left(\rho,T\right)\nu_{ei}^{f_3\left(\rho,T\right)}\right)}+D_0 \\
     &=D_{\text{CIF}}+D_0
 \end{split} \end{equation}
 
\begin{figure}
     \centering
           \includegraphics[width=0.4\textwidth]{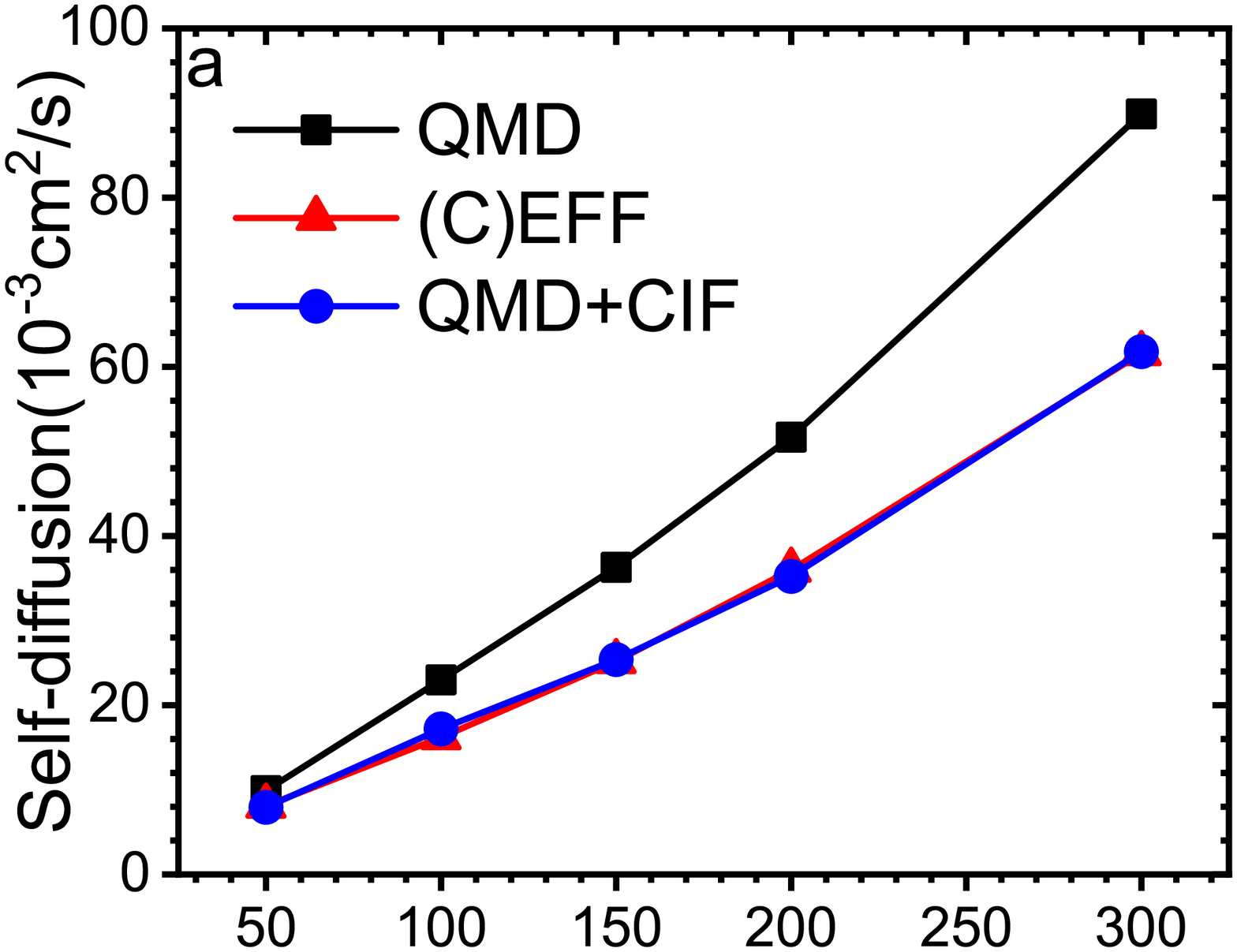}   \\
           \includegraphics[width=0.4\textwidth]{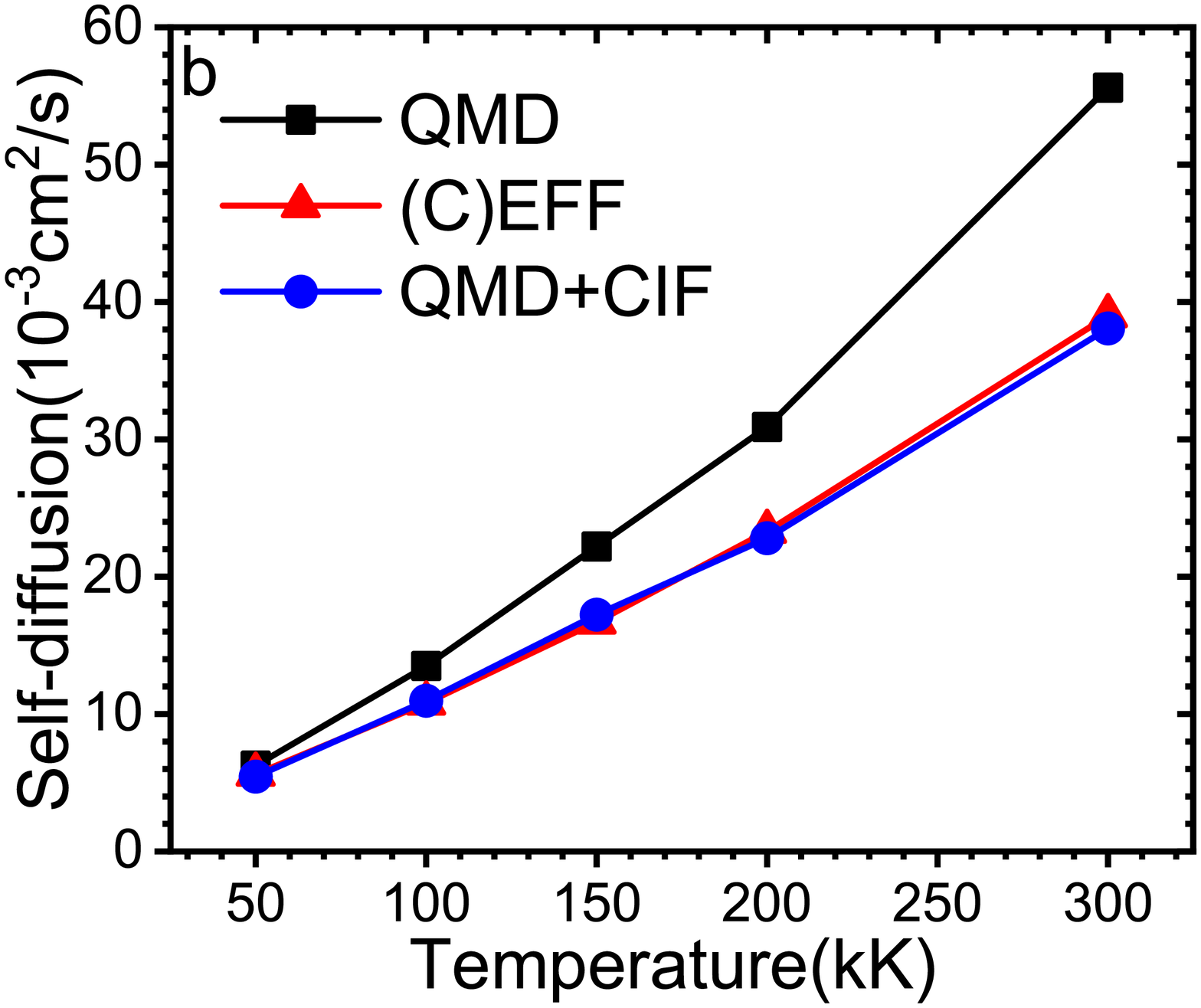}
     \caption{\label{fig:verify}Self-diffusion coefficients of H calculated by different methods at $5\text{g/cm}^3$ (a) and $10\text{g/cm}^3$ (b). The solid black squares and red triangles represent the QMD and (C)EFF results, respectively. The CIF model is used to correct the QMD results as the consideration of non-adiabatic effect. The results are represented by the blue circles.}
 \end{figure}

For $D_0$, plenty of models have been developed to study on it, such as QMD and OFMD which are based on BO approximation. In this paper, the diffusion coefficient including non-adiabatic effect has been calculated using (C)EFF method. As the collision frequency is a small term, the equation can be simplified as $D_{\text{CIF}}=D-D_0=f\left(\rho,T\right)\nu_{ei}^{f'\left(\rho,T\right)}$. $D$ and $D_0$ can be obtained from (C)EFF and QMD simulations, respectively. We develop an empirical fitting function from the available data as the assessment of electron-ion collisions induced the decrease of ionic diffusions
\begin{equation}
     D_{\text{CIF}}=\frac{\nu_{ei}^{0.25}}{a\rho/T^{3/2}+b\rho+c/T^{3/2}+d}
\end{equation}
where the fitting coefficient $a=-8.942\times10^{-3},b=1.585\times10^{-3},c=6.849$, and $d=-4.195$.The corrected QMD results by the CIF model, which are shown in Fig. \ref{fig:verify}, agree well with (C)EFF simulations. To verify the accuracy of the fitting function, we calculate self-diffusion of H and He at some other temperatures and densities. The results are listed in Table \ref{tab:verify}.
 
\begin{table*}
     \caption{\label{tab:verify}The self-diffusion coefficients calculated by QMD, QLMD and (C)EFF models. The QMD results corrected by the CIF model is also listed in the table.}
     \begin{ruledtabular}
     \begin{tabular}{ccccccc}
         species & density$(\text{g/cm}^3)$ & temperature(K) & $D_{\text{QMD}}(\text{cm}^2/\text{s})$ & $D_{\text{QLMD}}(\text{cm}^2/\text{s})$ & $D_{\text{(C)EFF}}(\text{cm}^2/\text{s})$ & $D_{\text{QMD}+\text{CIF}}(\text{cm}^2/\text{s})$ \\
        \hline
        H  & 8 & 100000 & 0.0155 & 0.0133 & 0.0122 & 0.0123  \\
        H  & 8 & 200000 & 0.0386 & 0.029 & 0.0264 & 0.0284  \\
        H  & 15 & 200000 & 0.0237 & 0.0181 & 0.0182 & 0.0183  \\
        H  & 15 & 300000 & 0.0396 & 0.0289 & 0.03 & 0.0278   \\
        He  & 10 & 100000 & 0.00757 & 0.0066 & 0.00598 & 0.00596   \\
        He  & 10 & 200000 & 0.0181 & 0.016 & 0.0108 & 0.0128   \\
     \end{tabular}
     \end{ruledtabular}
 \end{table*} 

In Table \ref{tab:verify}, the ionic self-diffusion coefficients obtained from the QMD model can be modified by adding the CIF factor as the compensation of electron-ion collisions. The results are in good agreement with the (C)EFF and QLMD results, showing that our CIF modification can be applied to warm dense matter. It needs to be emphasized that the CIF modification is independent of other models, therefore, any model based on the adiabatic framework can use it to offset the lost of the electron-ion collisions.
 
 \section{\label{sec:five}Comparision with analytical models}

The expensive computational costs of first principles simulations make it difficult to apply online or generate large amount of data. On the contrary, some analytical models based on numerical simulations have been proposed, supplying promising approaches to the establishment of database. However, the accuracy of these models should be examined when applied to WDM \cite{ZGLi2016}. In this section, we use QMD and our modified QMD results as benchmark trying to find an efficient and accurate model to acquire transport parameters.

We firstly compare our results with the Yukawa one-component plasma (YOCP) model, which is a development version of OCP model \cite{Daligault2006,Daligault2009}. In the YOCP model, the electron screening is included to modify the bare Coulomb interactions \cite{Hamaguchi1997,Murillo2000,Daligault2012}. The interaction between ions is replaced by the Yukawa potential
\begin{equation}
    u\left(r\right)=q^2e^{-{\kappa}r}/r
\end{equation}
where $\kappa$ is the inverse screening length. All properties of the YOCP model are dependent on the inverse screening length $\kappa$ and the coupling parameter $\Gamma$. Daligault has applied the model in a wide range of $\kappa$ and over the entire fluid region \cite{Daligault2012b}. In the gas-like small coupling regime, the reduced self-diffusion coefficients model can be extended from the Chapman-Spitzer results as \cite{Daligault2012b}
 \begin{equation}
     D^*\left(\kappa,\Gamma\right)=\sqrt{\frac{\pi}{3}}\frac{1}{\alpha\left(\kappa\right)}\frac{1}{\Gamma^{5/2}\ln {\Lambda\left(\kappa,\Gamma\right)}}
 \end{equation}
The generalized Coulomb logarithm $\ln {\Lambda\left(\kappa,\Gamma\right)}$ is expressed as
 \begin{equation}
     \ln {\Lambda\left(\kappa,\Gamma\right)}=\ln{\left(1+B\left(\kappa\right)\frac{\lambda_D}{b_c}\right)}=\ln{\left(1+\frac{B\left(\kappa\right)}{\sqrt{3}\Gamma^{3/2}}\right)}
 \end{equation}
where $\lambda_D$ is the Debye length $\lambda_D=\sqrt{4{\pi}q^2n/k_BT}$ and $b_c$ is the classical distance of closest approach $b_c=Zq^2/k_BT$. $\alpha\left(\kappa\right)$ and $B\left(\kappa\right)$ are fitting parameters dependent on $\kappa$ only
 \begin{gather}
     \alpha\left(\kappa\right)=\sqrt{\frac{3}{\pi}}\frac{1}{a_0+a_1\kappa^{a_2}} \\
     B\left(\kappa\right)=b_0+b_1\textrm{erf}\left(b_2\kappa^{b_3}\right)
 \end{gather}
with $a_0=1.559773, a_1=1.10941, a_2=1.36909, b_0=2.20689, b_1=1.351594, b_2=1.57138$, and $b_3=3.34187$.  

\begin{figure}
     \centering
           \includegraphics[width=0.4\textwidth]{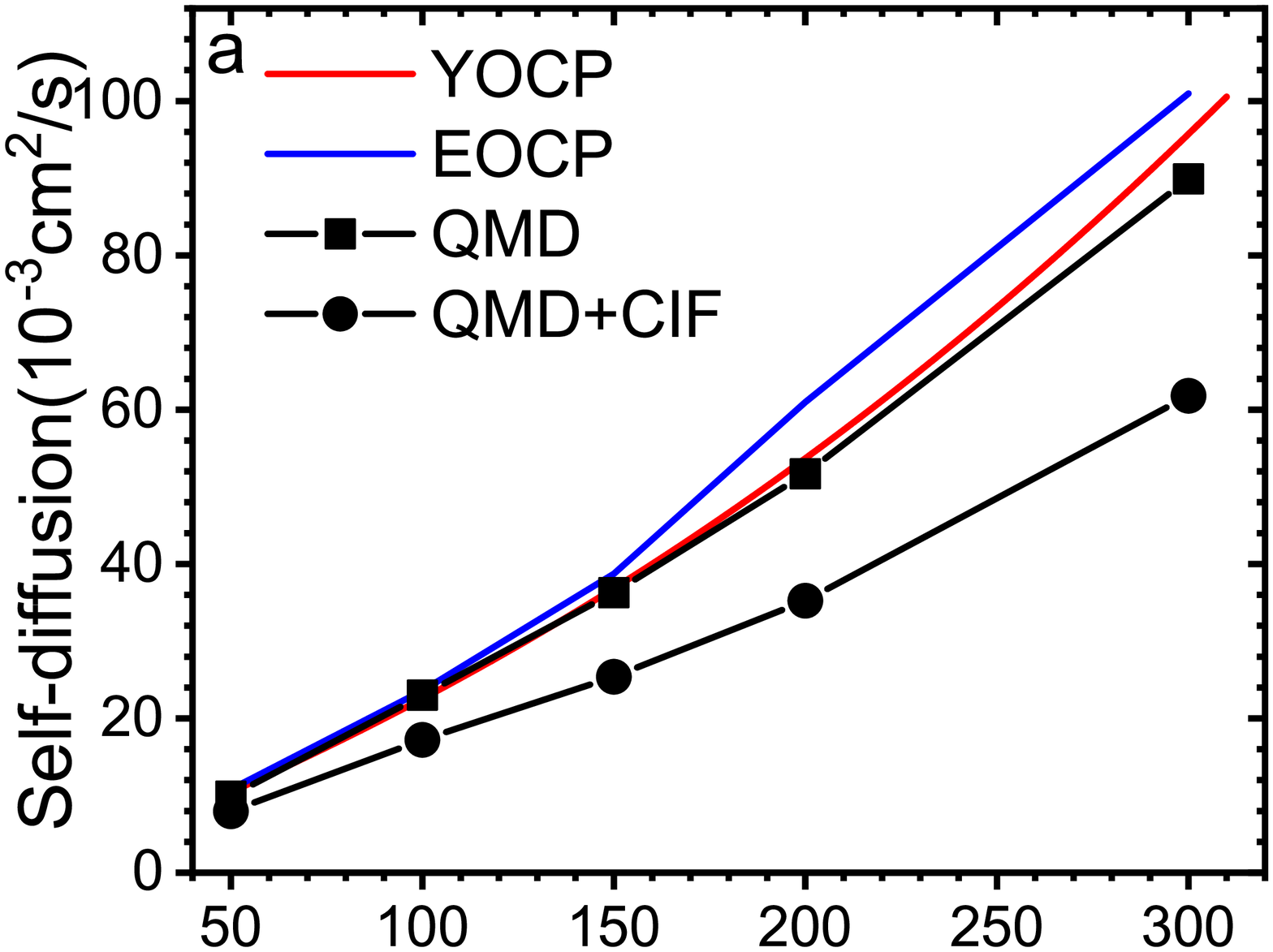}
           \includegraphics[width=0.4\textwidth]{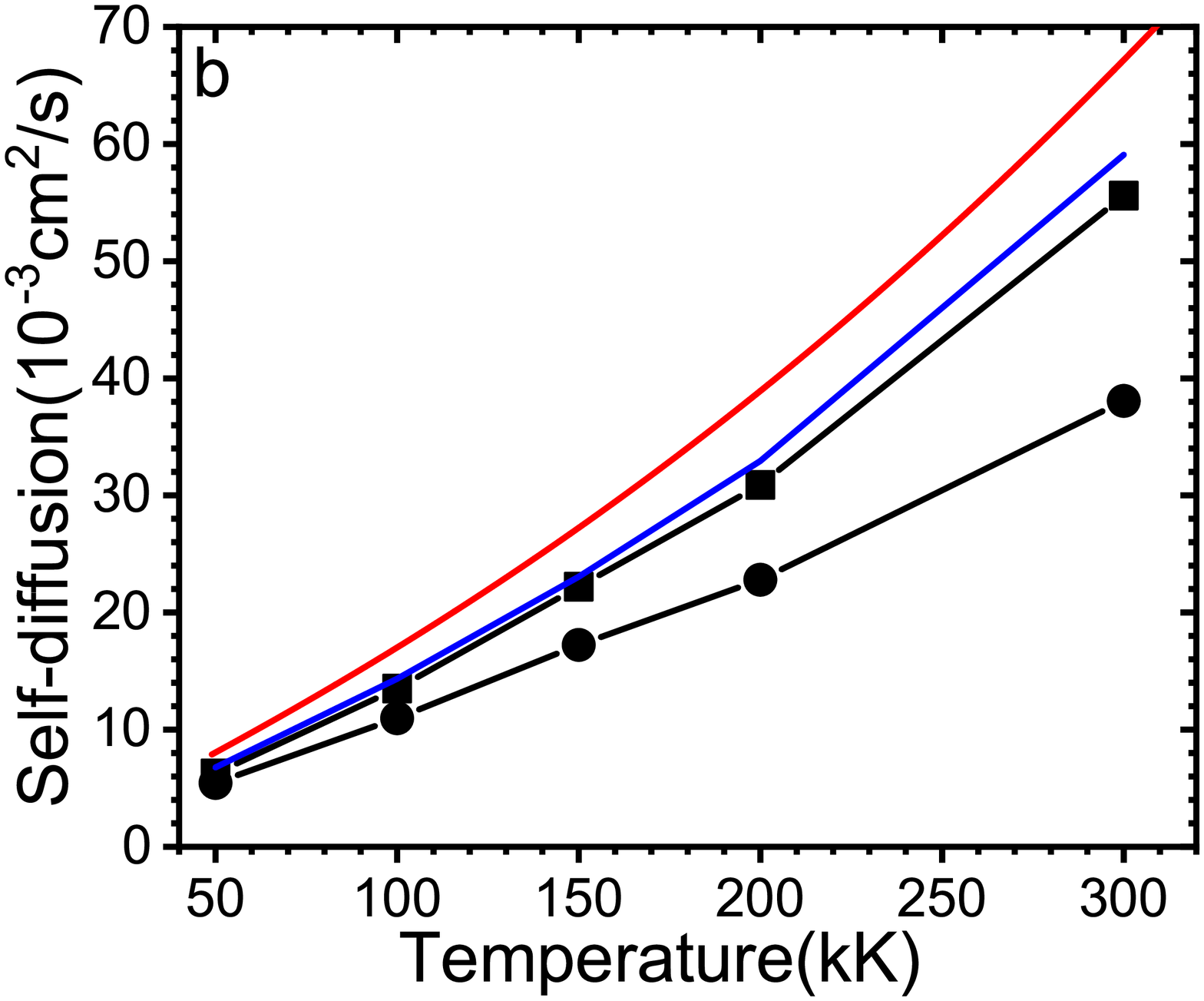}
     \caption{\label{fig:diff_models}Comparison of QMD and modified QMD simulations with different analytical models for self-diffusion coefficients of warm dense H at $5\text{g/cm}^3$ (a) and $10\text{g/cm}^3$ (b). The black solid squares and circles represent the results calculated by QMD and QMD with CIF correction, respectively. The red and blue solid lines are the results of the YOCP model\cite{Daligault2012b} and EOCP model\cite{Clerouin2016}. The effective coupling parameters of the EOCP model are obtained from the RDFs of QMD.}
 \end{figure}

Here, we use the Thomas-Fermi (TF) length \cite{Glenzer2009} to estimate the screening of electrons, so that
\begin{equation}
     \kappa=\frac{1}{\lambda_{TF}}=\frac{1}{\left(\pi/12Z\right)^{1/3}\sqrt{r_i}}
\end{equation}         
where $Z$ is the ionic charge, and $r_i$ is the Wigner-Seitz radius defined as $r_i=\left(3/\left(4{\pi}n_i\right)\right)^{1/3}$.The self-diffusion coefficient is obtained according to $D=D^*{\omega}a^2$, and $\omega=\left(4{\pi}n_iZ^{*2}e^2/m_i\right)^{1/2}$, where $Z^*$ is the average ionization degree which is calculated by the AA model \cite{YHou2006}. The comparision between the results of different models are shown in Fig.~\ref{fig:diff_models}. The QMD results and the CIF modification of QMD's are also shown as benchmarks.

As shown in Fig.~\ref{fig:diff_models}, for warm dense hydrogen at the density of $5\text{g/cm}^3$, the YOCP model can excellently reproduce results from the QMD simulations. However, at higher density, the YOCP model overestimates the diffusions compared with QMD results. This is because TF length overestimates the screening. With the density increase, electronic charges are excluded from dense ions that ionic repulsion becomes stronger at short range \cite{HYSun2017,Daligault2016}. And the correlation of ions becomes weaker leading to lower diffusions which is not considered in the TF model. To modify the YOCP model, we should adjust screening length artificially \cite{Zerah1992,Clerouin2001}. There is another scheme to deal with larger ionic coupling systems, in which we can reduce effective volumes of particles so that the collision frequency can be increased and the ionic transportation can be dragged or dissipative. The model has successfully improved the transport properties of strongly coupled plasmas in the range $1\leq\Gamma\leq30$ \cite{Daligault2016,Baalrud2015,Baalrud2013}.  

As another method, the EOCP model has a better description for all density and temperature range we studied with the QMD simulations as shown in Fig.~\ref{fig:diff_models}. In EOCP model, The effective coupling parameter $\Gamma_e$ and ionization $Q_e$ are introduced as the correction of the OCP model to reproduced the static structures of the OFMD's\cite{Clerouin2013}, the model also works well on transports properties such as diffusion and viscosity \cite{Arnault2013,Clerouin2016}. In this paper, we set $\Gamma_e$ by the procedure developed by Ott \emph{et al} \cite{Ott2014} as
 \begin{equation}
     \Gamma_e=1.238\exp{\left(1.575r^3_{1/2}\right)}-0.931,\quad\left(r_{1/2}<1.3\right)
 \end{equation}
where $r_{1/2}$ is obtained from the RDFs $g\left(r\right)$ at $g\left(r\right)=0.5$, The distance is expressed in the Wigner-Seitz radius unit. The effective average charge $Q_e$ is defined as $Q_e=\sqrt{\Gamma_eak_BT}/e$. We use the RDFs of QMD's as the input of EOCP model, the results agree well with those extracted from long time MD simulations, especially when temperature is low. Compared to the YOCP model, the EOCP model results give a more reasonable description of ionic diffusions. The EOCP model extracts the information directly from the static structure of the system. While the accuracy of the YOCP model depends on the selection of the particle interactions, which should be modeled \emph{a priori}.    

However, neither model agrees well with modified QMD results. This can be attributed to the loss of the non-adiabatic effect of the two models. Both the YOCP model and the EOCP model calculate self-diffusion coefficients based on the static potential, and the dynamic electron ion collisions can not be considered in it. This reminds us to pay attention to the instantaneous dynamic collisions effect when doing MD simulations. For application, we can use the CIF model to modify the EOCP model, which is a cheaper way to obtain the self-diffusion data including the non-adiabatic effect. 

\section{\label{sec:six}CONCLUSION}

We have performed QMD, OFMD, and (C)EFF simulations to determine the RDFs and the ionic self-diffusion coefficients of warm dense hydrogen at the densities of $5\text{g/cm}^3$ and $10\text{g/cm}^3$ and temperatures from 50kK to 300kK. The results from (C)EFF-MD method are carefully compared with the results from QMD/OFMD methods based on the BO approximation. In EFF method, the static properties are insensitive to electron-ion collisions, however, the diffusion of ions decreases significantly with the increase of electron-ion collisions. The ionic diffusion coefficients calculated from (C)EFF agree well with the QLMD results, but largely differ from QMD or OFMD simulations, revealing key role of electron-ion collisions in warm dense hydrogen. Most importantly, we proposed a new analytical model which introduce the electron-ion collisions induced friction (CIF) effects, constructing a formula to calculate self-diffusion coefficients without doing non-adiabatic simulations. The CIF model has been verified to be valid over a wider range of temperature, density and materials. However, since the CIF model is derived from the fitting of simulation results, whether it can be applied for more complex elements should be verified further. We also show the results from analytical models of YOCP and EOCP. Based on the static information, EOCP model reproduces QMD simulations better. However, neither of the two models considers the dynamic electron-ion collisions effect. We propose to use the CIF model to modify the EOCP results as a preferred scheme to calculate self-diffusion coefficients.
 
 \section{ACKNOWLEDGMENTS}

The authors thank Dr. Zhiguo Li for his helpful discussion. This work was supported by the Science Challenge Project under Grant No. TZ2016001, the National Key R\&D Program of China under Grant No. 2017YFA0403200, the National Natural Science Foundation of China under Grant Nos. 11774429 and 11874424, the NSAF under Grant No. U1830206. All calculations were carried out at the Research Center of Supercomputing Application at NUDT.

 \section{Data Available}

The data that support the findings of this study are available within the article.

\bibliography{references}

\end{document}